\documentstyle[12pt,epsf]{article}
\newcommand{\be}{\begin{equation}}
\newcommand{\ee}{\end{equation}}
\newcommand{\ba}{\begin{eqnarray}}
\newcommand{\ea}{\end{eqnarray}}
\newcommand{\n}{\nonumber\\}
\begin{document}
\title{Chern-Simons action in noncommutative space}
\author{ N.~Grandi\footnote{grandi@venus.fisica.unlp.edu.ar}
~ and G.A.~Silva\footnote{silva@venus.fisica.unlp.edu.ar} \\
  \normalsize {\it Departamento de F\'\i sica, Universidad
  Nacional de La Plata}\\
  {\normalsize\it C.C. 67, 1900 La Plata, Argentina}}
\date{\hfill}
 \maketitle
\begin{abstract}
We derive the noncommutative Chern-Simons action induced by Dirac
fermions coupled to a background gauge field, for the fundamental,
antifundamental, and the adjoint representation. We discuss
properties of the noncommutative Chern-Simons action showing in
particular that the Seiberg-Witten formula maps it into the
standard commutative Chern-Simons action.
\end{abstract}

Recent results  in noncommutative geometry and string theory
\cite{CDS}-\cite{sw}, revealed  the interest on its own right for
studying different field theories like Yang-Mills, $\lambda
\phi^4$, QED, Chern-Simons, Wess-Zumino theories and
two-dimen\-sio\-nal models, in noncommutative space
\cite{S}-\cite{CG1}. In this respect, it is the purpose of this
paper to analyse different aspects of the noncommutative
Chern-Simons (CS) action. First,  we discuss how the parity
anomaly in a $2+1$ massive fermion model induces a Chern-Simons
term (as originally observed in \cite{chu} for the massless case).
Then, we discuss relevant properties of the non-commutative CS
action, its relation with the chiral Wess-Zumino-Witten model and
its dependence on the noncommutative parameter $\theta_{\mu\nu}$.

Let us start by establishing our conventions. The $*$-product for
fields is defined by
\be
 (\hat f * \hat g)(x) =
 \left.
 e^{\frac i 2{\theta_{\mu\nu}}\partial_{\xi_\mu}
 \partial_{\zeta_\nu}}\hat f(x+\xi) \hat g(x+\zeta)
 \right\vert_{\xi=\zeta=0}
 \label{star}
\ee
and the Moyal brackets as
\be
 \left\{
 \hat f(x),\hat g(x)
 \right\}=\hat f(x)*\hat g(x)-\hat g(x)*\hat f(x)
 \label{moyal}
\ee
We indicate with a hat  functions which have to be multiplied
using the $*$-product. The $U(N)$ gauge-group elements are defined
by
\be
 \hat g(x) = e_*^{\,i \hat \alpha(x)} = 1 + i \,\hat \alpha(x)-
 \frac {1} 2 \,
 \hat \alpha(x)*\hat \alpha(x)
 +\cdots
\ee
where $\hat \alpha(x)$ is a Lie-algebra valued function of
space-time. Gauge fields in the Lie algebras of $U(N)$ transform
according to
\be
 \hat A_\mu(x)  \to \hat g(x)*\hat A_\mu(x)*\hat g^{-1}(x)- \frac i
 e \hat g(x)*\partial_\mu\hat g^{-1}(x) \label{gg}
\ee
with the field strength defined as
\be
 \hat F_{\mu\nu} = \partial_\mu\hat A_\nu-\partial_\nu\hat A_\mu +i
 e \left\{\hat A_\mu,\hat A_\nu \right\}~.
 \label{def}
\ee
When acting over fermion fields $\hat \psi$, even in the $U(1)$
case, there are three possible representations of the gauge group
action
\be
 \hat \psi(x)
 \to\left\{
 \begin{array}{ll}
 \hat g(x)*\hat \psi(x)&${\rm fundamental representation} $``f"$ $\\
 \hat \psi(x)*\hat g^{-1}(x)&${\rm anti-fundamental representation}
 $``\bar f"$ $\\
 \hat g(x)*\hat \psi(x)*\hat g^{-1}(x)&${\rm adjoint representation}
 $``ad"$ $
 \end{array}
 \right.
\ee
Accordingly, the covariant derivative acting on $\psi$ is defined
as
\be
 \hat D_\mu \hat\psi(\tau,x) =
 \left\{
 \begin{array}{ll}
 \partial_\mu \hat\psi + i e \,{\hat A}_\mu\!*\hat\psi &
 $ $``f" $ $
 \\
 \partial_\mu \hat\psi - i e\, \hat\psi* \hat A_\mu &
 $ $``\bar f"$ $
 \\
 \partial_\mu \hat\psi + i e\,\{\hat A_\mu,\hat\psi\} &
 $ $``ad"$ $
 \end{array}
 \right.~.
 \label{defi}
\ee
We write the action for massive fermions, coupled to a gauge
field, in $2+1$ non-commutative space as
\be
 S(\hat A;m) \;=\;  \int d^3 x \; {\bar {\hat\psi}}({x})* ( i\not
 \!\! \hat D -m ) \hat \psi({x}) \; .
 \label{fund}
\ee
and define the effective action $\Gamma(\hat A;m)$ trough
\be
e^{i\Gamma(\hat A; m)} =
Z(\hat A; m)
=
\int{\cal D}{\hat \psi}{\cal D}{\bar{\hat \psi}}
\,e^{iS(\hat A; m)}
\label{Z}
\ee

\subsection*{Induced Chern-Simons term}
Before studying some specific properties of non-commutative
Chern-Simons action, let us describe how a parity violating  Chern-Simons
term is induced by fluctuations of massive non-commutative
fermions fields, exactly as it happens
in the commutative case
\cite{redlich}. We shall just concentrate in the parity
odd part of the effective action $\Gamma_{odd}$, thus disregarding
parity conserving contributions.

\subsubsection*{Fundamental and anti-fundamental representations}

The calculation of the effective action for fermions in the
fundamental and the anti-fundamental representations gives the
same answer. We shall describe first the case of the fundamental
representation. As in the original calculation in \cite{redlich},
one obtains the contribution to $\Gamma_{odd}(\hat A; m)$  from
the vacuum polarization and the triangle graphs
\ba
 \label{sss}
 i \Gamma_{odd} [\hat A;m] &=& \left. \left(
 \frac{1}{2} {\rm Tr} \int \frac{ d^3 p}{(2\pi)^3}  {\hat A}_\mu
 (p) \,\Pi^{\mu\nu}(p;m)\, {\hat A}_\nu (-p)  + \right. \right. \n
 && \!\!\!\!\!\!\!\!\!\! \left. \left. +\frac{1}{3}{\rm Tr}
 \int \frac{ d^3 p}{(2\pi)^3}\frac{ d^3
 q}{(2\pi)^3}\Gamma^{\mu\nu\rho}(p,q;m)\,{\hat A}_\mu(p) {\hat
 A}_\nu(q) {\hat A}_\rho (-p-q) \right) \right\vert_{odd}
 \n
 \label{unoo}
\ea
Here Tr represents the trace over the $U(N)$ algebra generators,
with
\be
\label{111}
 \Pi^{\mu \nu}(p;m) = -e^2
\int \frac { d^3 k}{(2\pi)^3} {\rm tr} \left[ \gamma^\mu
\frac{\slash\!\!\!k -m}{k^2 -m^2}\gamma^\nu \frac{\slash\!\!\!k
+\slash\!\!\!\!p - m} {(k+p)^2 -m^2} \right] \ee
\ba
 \label{222}
 \Gamma^{\mu\nu\rho}(p,q;m) \!\!\!&=&\!\!\!
 e^3\,\exp(- \frac{i}{2} p_\lambda \theta^{\lambda\delta} q_\delta
 ) \int \!\frac{ d^3 k}{(2\pi)^3} {\rm tr}\! \left[\gamma^\mu
 \frac{(\slash\!\!\!k - m)}{k^2 -m^2} \gamma^\nu \frac{
 (\slash\!\!\!k-\slash\!\!\!q -m)}{(k-q)^2 -m^2}\times \right.
 \n
 && \;\;\;\; \left. \times \gamma^\rho
 \frac{(\slash\!\!\!k +\slash\!\!\!p -m)}{(k+p)^2-m^2} \right]
\ea
As first observed in \cite{chu} for massless fermions, there are
no nonplanar contributions to the parity odd sector of the
effective action, the only modification arising from
noncommuativity is the $\theta$-dependent phase factor in
$\Gamma^{\mu\nu\rho}$, associated to external legs in the cubic
term, which is nothing but the star product in configuration
space. The result for $\Gamma_{odd}(\hat A;m)$ is analogous to the
commutative one except that the star $*$-product replaces the
ordinary product.

Regularization of the divergent integrals (\ref{111}) and
(\ref{222}) can be achieved by introducing in the original action
(\ref{fund}), bosonic-spinor Pauli-Villars fields with mass $M$.
These fields  give rise to additional diagrams, identical to those
of eq.(\ref{unoo}), except that the regulating mass $M$ appears in
place of the physical mass $m$. Since we are interested in the
parity violating part of the effective action, we keep only the
parity-odd terms in (\ref{111}) and (\ref{222}) (and in the
corresponding regulator field graphs). To leading order in
$\partial/m$, the gauge-invariant parity violating part of the
effective action is, for the fundamental representation, given by
\ba
 \Gamma_{odd}^f(\hat A,m)
 &=& \frac{1}{2}
 \left( \frac{m}{|m|}+ \frac{M}{|M|} \right) {\hat S}_{CS}(\hat A)
 +O(\partial^2/m^2)
 \n
 &=& \pm  {\hat S}_{CS}(\hat A)
 +O(\partial^2/m^2)
 \label{mM}
\ea
with
\be
{\hat S}_{CS}(\hat A) = \frac{e^2}{4\pi}\int d^3x\, \epsilon^{\mu\nu\rho} {\rm
tr} \left( \hat A_\mu *
\partial_\nu \hat A_\rho\right.\nonumber\\
\left.+  \frac{ 2ie}{ 3} \hat A_\mu* \hat A_\nu* \hat A_\rho
\right)
\label{ec}
 \ee
As it is well known, the relative sign of the fermion and
regulator contributions depends on the choice of the Pauli-Villars
regulating Lagrangian (of course the divergent parts should cancel
out independently of this choice). In the first line of (\ref{mM})
we have made a choice such that the two contributions add to give
the known Chern-Simons result of the second line. Note that even
in the Abelian case, the Chern-Simons action contains a cubic term
(analogous to that arising in the ordinary non-Abelian case).

As expected, the effective action is gauge invariant even under
large gauge transformations. This is due to the fact that we have
taken into account both parity violating sources: that originated
in the (parity non-invariant) fermion mass term and that related
to the regularization prescription (which requires the
introduction of the mass $M$).

As advanced, the parity odd part of the effective action for
fermions in the anti-fundamental representation gives the same
answer. There is a change of sign $e \to -e$ on each vertex,
compensated by a change in the momenta dependence of propagators
due to the different ordering of fields in the $f$ and $\bar f$
covariant derivatives (see (\ref{defi})).

\subsubsection*{Adjoint representation}

The diagrams contributing to $\Pi^{\mu \nu}$ in the adjoint
representation are shown in Fig. 1.


\vspace{-0.1 cm} \epsfxsize=4.4in \epsffile[1 500 480 750]{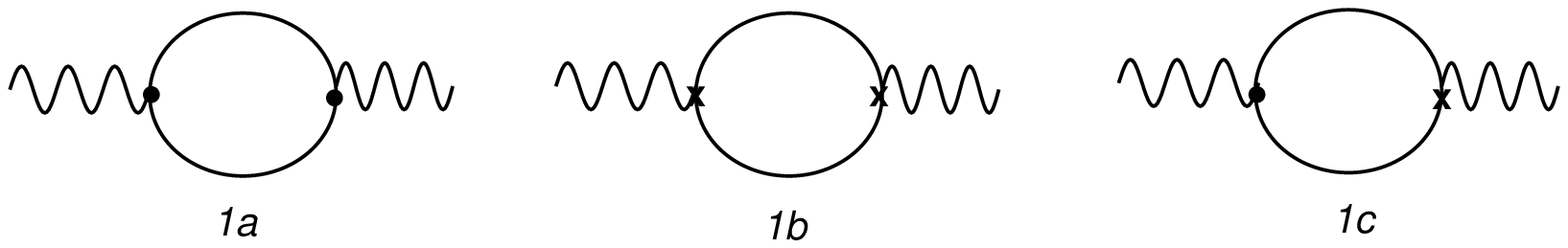}
\vspace{-2.6 cm}

\centerline{Figure 1}

\noindent{\small
{\it Diagrams contributing to $\Pi^{\mu \nu}$ for
fermions in the adjoint representation. The
dotted vertex coincides with the coupling of $A_\mu$ with
fermions in the
fundamental representation, the cross with that for
fermions in the
anti-fundamental.}}


\vspace{0.5 cm}

Planar diagrams $1a$ and $1b$ coincide with those arising in the
fundamental and the anti-fundamental representation, thus giving,
each one, the previously computed answer (\ref{mM}). Concerning
the non-planar diagram $1c$, the resulting contribution is given
by
\be
 \label{113}
 \Pi^{\mu \nu}_{1c}(p;m) =  e^2 \int \frac { d^3 k}
 {(2\pi)^3}\exp(-i  p_\lambda \theta^{\lambda\delta } k_\delta )
 {\rm tr}
 \left[ \gamma^\mu \frac{\slash\!\!\!k -\slash\!\!\!\!p -m}
 {(k-p)^2 -m^2}
 \gamma^\nu \frac{\slash\!\!\!k -m}{k^2 -m^2} \right]
\ee
The parity odd part of the above expression is
\ba
 \left.\Pi^{\mu \nu}_{1c}(p;m)\right\vert_{odd}
 &=& -2 e^2m \epsilon^{\mu\nu\rho}\,ip_\rho
 \int \frac { d^3 k}{(2\pi)^3} \frac{\exp(-i p_\lambda
 \theta^{\lambda\delta}
 k_\delta )}{(k^2 -m^2)((k-p)^2 -m^2)}
 \n
 &=&- 2\frac m {|m|} e^2
 \epsilon^{\mu\nu\rho}\,ip_\rho
 \int \frac { d^3 q}{(2\pi)^3} \frac{\exp(-i|m| p_\lambda
 \theta^{\lambda\delta} q_\delta )}{(q^2 -1)((q-\frac p{|m|})^2-1)}
 \label{114}
\ea
where we have written $k_\mu = |m|q_\mu$. As in the previous
section, one should add the regulator contribution.

We are interested in the leading term in a derivative expansion of
the effective action. In the ordinary (commutative) case, this
amounts to make an expansion in powers of the unique available
dimensionless variable,  $p/m$. In the noncommutative case, where
one has, apart from the fermion mass, the dimensionfull parameter
$\theta$, one can construct a second independent dimensionless
variable, $m|p\theta|$. Let us first expand (\ref{114}) to first
order in $p/m$,
\ba
 \left.\Pi^{\mu \nu}_{1c}(p;m)\right\vert_{odd} &=& -\frac m
 {|m|} 2i e^2
  \epsilon^{\mu\nu\rho}\,p_\rho
 \int \frac { d^3 q}{(2\pi)^3} \frac {\exp(-i \vert m\vert
 p_\lambda \theta^{\lambda\delta} q_\delta )} {(q^2 -1)^2}
 \n
 && - \frac M {|M|} 2i e^2
  \epsilon^{\mu\nu\rho}p_\rho
 \int \frac { d^3 q}{(2\pi)^3} \frac {\exp(-i \vert M\vert
 p_\lambda \theta^{\lambda\delta} q_\delta )} {(q^2 -1)^2}
 \label{psm}
\ea
here, the regulator contribution has been explicitly written.
Concerning the expansion in powers of the second dimensionless
parameter $m|p\theta|$, let us note that, since $m$ is finite,
first order in $m|p\theta|$ should be kept in the first term of
(\ref{psm}). This gives the same contribution to the effective
action as the $1a$ and $1b$ graphs. For the second term in
(\ref{psm}), the $M\to \infty$ limit must be taken, then, the
oscillating factor makes the integral vanish \cite{chu}. Finally
one gets
\ba
 \left.\Pi^{\mu\nu}_{1c}(p;m)\right\vert_{odd}
 &=& -i\frac m {|m|}\frac{e^2}{4\pi}
  \epsilon^{\mu\nu\rho}ip_\rho~,
\label{ppsm}
\ea
so that the complete quadratic $\Pi_{\mu\nu}$ for the
adjoint representation is then given by
\be
 \Pi^{\mu\nu}_{adj}(p;m)
 =\Pi^{\mu\nu}_{1a}(p;m) + \Pi^{\mu\nu}_{1b}(p;m)+
 2\Pi^{\mu\nu}_{1c}(p;m)
 =\frac{e^2}{2\pi}\epsilon_{\mu\nu\rho}p_\rho
 \frac M {|M|}
 \label{ad}
\ee

Note that the whole contribution to $\Pi^{\mu\nu}$ in the adjoint
comes from the regulating fields. This accounts for the quadratic
part of the CS induced action. Concerning the cubic term, it can
be either explicitly computed or adjusted so as to achieve
gauge-invariance. In anycase,  the result for the the parity
violating effective action for fermions in the adjoint is, to
leading order in $\partial$,
\be
\Gamma_{odd}^{ad}(\hat A,m) =\pm
 \hat S_{CS}(\hat A) +  O(\partial^2)~.
 \label{com}
\ee
As before, the result is gauge invariant even under large gauge
transformations.

 It should be stressed that (\ref{com}) gives a non-trivial effective
action even in the   $\theta \to 0$ limit, in which fermions in
the adjoint decouple from the gauge field. As observed in other
cases \cite{MRS}-\cite{RS}, \cite{FI}, \cite{MS2}, this is due to
the fact that this limit does not commute with that of the
regulator $M \to \infty$.

\subsection*{The connection between noncommutative CS and chiral
WZW theories}

As it is well-known, the (ordinary) CS theory can be related with
the chiral WZW model following different approaches
\cite{Wi}-\cite{Eal}. Here, we shall discuss how such a connection
can be established  in the noncommutative case.

Consider the  action
\be
 \hat S_{CS}[\hat A_0,\hat A_i]=\frac { e^2} {4\pi}{\rm Tr} \int_{{\cal M}}
 d^3x\, \epsilon_{ij}\left({\hat A}_0* {\hat F}_{ij}+ \dot {\hat
 A}_i *{\hat A}_j\right)~,
 \label{wzw1}
\ee
which differs from the  CS action
(\ref{ec}) by a surface term. Of course,  when ${\cal M}$ has no
boundary, such surface term is irrelevant.  However, in what
follows we choose as manifold ${\cal M} = \Sigma \times R$ with
$\Sigma$ a two-dimensional manifold. We shall take eq.(\ref{wzw1})
as the starting point for quantization of the $2+1$ theory and
follow the steps described in \cite{MSe}-\cite{Eal} in their
original derivation of the connection.

Expression  (\ref{wzw1}) can be rewritten as
\be
 \hat S_{CS}[\hat A_0,\hat A_i]= \frac { e^2} {4\pi}{\rm Tr} \int_{{\cal
 M}} d^3x\, \epsilon_{ij}\left({\hat A}_0 {\hat F}_{ij} + \dot {\hat
 A}_i *{\hat A}_j\right) + \frac { e^2} {4\pi}{\rm Tr} \int_{\partial {\cal
 M}} dS_{\mu}\Lambda^{\mu}
 \label{wzw2}
\ee
with
\be
 \Lambda^\mu = \epsilon_{ij}\sum_{n=1}^{\infty}\frac 1 {n!}
 \left(\frac i 2\right)^n \theta^{\mu
 \nu_1}\theta^{\mu_2\nu_2}...\theta^{\mu_n\nu_n}
 \partial_{\mu_2}...\partial_{\mu_n}{\hat A}_0\,\partial_{\nu_1}
 \partial_{\nu_2}...\partial_{\nu_n}{\hat F}_{ij}
\ee
 Using action (\ref{wzw2}),  the partition function for
 the noncommutative CS theory takes the form
\ba
 Z &=& \int {\cal D}\hat A_i {\cal D}\hat A_0
 \exp\left( \frac {i\kappa e^2} {4\pi}{\rm Tr}\int_{{\cal
 M}} d^3x\, \epsilon_{ij}\left({\hat A}_0 {\hat F}_{ij}+ \dot {\hat
 A}_i *{\hat A}_j\right)\right. \n
 &&\left.\;\;\;\;+\frac {i\kappa e^2} {4\pi}{\rm Tr}\int_{\partial{\cal
 M}}dS_{\mu}\Lambda^{\mu}\right)
 \label{z3}
\ea
where $\kappa$ is an integer. For interior points of ${\cal M}$,
$A_0$ acts as a Lagrange multiplier enforcing flatness of the
spatial components of the connection
\be
\hat F_{ij}(x) = 0 \;\;\;\;\;\;\;\;\; \forall x\in {\cal
M}-\partial{\cal M}
\label{flat}
\ee
By continuity, $\hat F_{ij}$ must also vanishes on the boundary.
The partition function takes then form
\be
 Z= \int {\cal D}\hat A_i \delta(\epsilon_{ij}\hat F_{ij})
 \exp\left(\frac {i\kappa e^2} {4\pi}{\rm Tr}\int_{\cal M}
 d^3x\,\epsilon_{ij}\dot{\hat A}_i* \hat A_j\right)
 \label{z22}
\ee

Let us  discuss the case where $\Sigma$ is the
disk.  Then  the solution of the flatness condition (\ref{flat}) is
$
\hat A_i = - \frac i e \,\hat g^{-1} *
\partial_i \hat g
$,
and one has reinserting it in (\ref{z22})
\be
Z = \int {\cal D}\hat g \exp\left(i\kappa\hat S_{CWZW}[\hat
g]\right) \label{zz3} \ee
where $\hat S_{CWZW}[\hat g]$ is the noncommutative, chiral WZW
action
%
\ba
 \hat S_{CWZW}[\hat g] &=& -\frac 1 {4\pi} {\rm Tr}\int_{\partial {\cal M}}
 d^2x (\hat g^{-1}*\partial_t \hat g)*(\hat
 g^{-1}*\partial_\varphi \hat g)
 \nonumber\\
 & &-\frac 1 {4\pi }{\rm Tr} \int_{{\cal M}} d^3x
 \,\epsilon_{ij}(\hat g^{-1} * \partial_i g)* ( \hat g^{-1} *
 \partial_t g)*( \hat g^{-1} * \partial_j g)
 \n
\ea
here $\varphi$ is a tangential coordinate which parametrize the
boundary of ${\cal M}_2$.

With this result in mind and taking into account the connection
between commutative and noncommutative  WZW models established in
\cite{MS2} through a Seiberg-Witten map, one can advance an
analogous connection for the CS theories. The situation can be
visulized in the following scheme
\be
 \begin{array}{ccccc}
    CWZW[{\hat g}] & \longleftrightarrow &
   \int \!d^3x\, ({\hat A}{\rm d} {\hat A}+\frac{2i}3 {\hat A}^3)  \\
   &&\\
   \updownarrow &   & \updownarrow\,\, ? \\
   &&\\
  CWZW[g] & \longleftrightarrow
  & \int\!d^3x\, ({ A}{\rm d} { A}+\frac{2i}3 { A}^3) \
 \end{array}
\ee
The next section is devoted to the study of this issue.

\subsubsection*{The Seiberg-Witten map}

A correspondence between commutative and noncommutative gauge
field theories can be defined  by the map \cite{sw}
\ba
\delta\hat A^\mu &=& \delta \theta^{\rho\sigma}
\frac \partial {\partial\theta^{\rho\sigma} } \hat
A_{\mu}(\theta)=
-\frac 1 4 \delta \theta^{\rho\sigma}
\left\{
\hat A_\rho ,
\partial_\sigma \hat A_\mu+\hat F_{\sigma\mu}
\right\}_+
\n
\delta\hat F_{\mu\nu}(\theta)&=&
\delta \theta^{\rho\sigma}
\frac \partial {\partial\theta^{\rho\sigma} } \hat
F_{\mu\nu}(\theta)
\nonumber\\
&=&
\frac 1 4 \delta\theta^{\rho\sigma}
\left(
2
\left\{
 \hat F_{\mu\rho},\hat F_{\nu\sigma}
\right\}_+
-
\left\{
\hat A_\rho,\hat D_\sigma\hat F_{\mu\nu}+\partial_\sigma\hat F_{\mu\nu}
\right\}_+
\right)
\label{swt}
\ea
For the case of noncommutative Yang-Mills action, this map leads
to a complicated non-polynomial commutative action. Remarkably, in
the Chern-Simons case, the action remains (up to surface terms)
invariant under the map (\ref{swt}). Let us write the
noncommutative Chern-Simons action in the form (\ref{ec})
\be
 \hat S_{CS}(\hat A)=
 \frac{e^2}{4\pi}\int_{{\cal M}} d^3x\, \epsilon^{\mu\nu\rho}
 \left(
 \hat A_\mu * \partial_\nu\hat A_\rho
 +
 \frac {2ie}{3}
 \hat A_\mu* \hat A_\nu* \hat A_\rho
 \right)
 \label{15}
\ee
where we choose for ${\cal M}$ either $R^3$ or $\Sigma\times R$
with $\Sigma$ a   manifold without boundary. Action (\ref{15}) can
be rewritten in the form
\ba
 \hat S_{CS}(\hat A) &=&\frac{e^2}{4\pi}\int_{{\cal M} }d^3x \,
 \epsilon_{ij} \left( {\hat  A}_0 {\hat F}_{ij} + \dot {\hat A}_i
 \hat  A_j \right)~.
 \label{F2}
\ea

In order to investigate the variation of this action under
Seiberg-Witten map, let us differentiate it with respect to
$\theta_{\mu\nu}$
\ba
 \frac{\partial {\hat S_{CS}(\hat A)}}{\partial\theta_{\mu\nu}}
 &=&\frac{e^2}{4\pi}\int_{{\cal M} } d^3x \,
 \epsilon_{ij}
 \frac \partial{\partial
 \theta_{\mu\nu}} \left( \hat A_0 \hat  F_{ij} +\dot A_i  A_j
 \right)\n
 &=&\frac{e^2}{4\pi}\int_{{\cal M} }d^3x \,
 \epsilon_{ij}
 \left( \frac {\partial
 \hat A_0}{\partial \theta_{\mu\nu}} \hat  F_{ij} + \hat A_0
 \frac{\partial \hat  F_{ij}}{\partial \theta_{\mu\nu}} + 2 \frac
 {\partial A_j}{\partial \theta_{\mu\nu}}\dot A_i \right)
\ea
Now, we can use (\ref{swt}) in order to rewrite the
$\theta$-derivatives. Keeping only the terms which are
antisymmetric with respect to the indices $\mu,\nu$ and $i,j$, we
get
\ba
 \frac{\partial {\hat S_{CS}(\hat A)}}{\partial\theta_{\mu\nu}}
 = 0 \;\;\;\;\;\;\;&\Rightarrow&\;\;\;\;\;\hat S_{CS}(\hat A)=
 S_{CS}(A)
 \label{a}
\ea
Here $S_{CS}(A)$ is the ordinary (commutative) CS action. It is
interesting to note that in the $U(1)$ case the SW map cancels out
the cubic term which is present in $\hat S_{CS}(\hat A)$.

In summary, we see that the SW transformation (\ref{swt}) maps the
noncommutative Chern-Simons action into the commutative one.

\subsection*{Conclusions}

We have computed the effective action for fermions in
noncommutative space, for different representations, showing that
a gauge invariant answer (even for large gauge transformations) is
obtained when regulator contributions are taken in account. In
particular, for the adjoint representation, the non-trivial gauge
invariant result (\ref{com}) is completely due to the regulator
fields, showing that the commutative $\theta\to 0$ limit does not
commute with the $M\to \infty$ limit.

We have shown that the noncommutative Chern-Simons action can be
related to the chiral noncommutative WZW model in the usual way.
It is important to note that for deriving this relation we needed
to define the Chern-Simons  theory from action (\ref{wzw1}), which
shows $A_0$ as a Lagrange multiplier enforcing the flatness
constraint (\ref{z3}). Finally, we showed that the Chern-Simons
action is mapped into the standard (commutative) action under the
Seiberg-Witten map (\ref{swt}).

\subsection*{Acknowledgements}

We are grateful to R.L. Pakman for collaboration at the early
stages of this work. We also acknowledge F.A. Schaposnik for many
useful conversations and encouragement and E.F. Moreno for
important remarks.

This work has been partly supported by CONICET and ANCPyT. The
authors are supported by CONICET fellowships.


\begin{thebibliography}{99}
\bibitem{CDS} A.Connes, M.R.~Douglas and A.S.~Schwarz,
JHEP {\bf 02} (1998) 003.
\bibitem{DH} M.R.~Douglas and C.~Hull, JHEP {\bf 02} (1998) 008.
\bibitem{sw} N.~Seiberg and E.~Witten, JHEP {\bf 09} (1999) 032.
\bibitem{S} M.M.~Sheikh-Jabbari, JHEP {\bf 06} (1999)015.
\bibitem{MRS}S.Minwalla, M.V~Raamsdonk and N.~Seiberg,
hep-th/9912072.
\bibitem{RS} M.V.~Raamsdonk and N.~Seiberg, JHEP {\bf 0003} (2000) 035.
\bibitem{MH} M.~Hayakawa, Phys.Lett. {\bf B478} (2000)  394;
 hep-th/9912167.
\bibitem{MST} A.~Matusis, L.~Susskind and N.Toumbas,
hep-th/0002075.
\bibitem{SST} N.~Seiberg, L.~Susskind and N.~Toumbas,
JHEP {\bf 0006}(2000) 044.
\bibitem{MS} E.~Moreno and F.A.~Schaposnik, JHEP {\bf 0003}  (2000) 032.
\bibitem{chu} C.S.~Chu, Nucl.Phys. {\bf B580} (2000) 352.
A.A.~Bichl, J.M.~Grimstrup, V.~Putz and M.Schweda, JHEP {\bf 0007}
(2000) 046.
\bibitem{DKL}L.~Dabrowski, T.~Krajewski and G.~Landi,
hep-th/0003099.
\bibitem{FI} K.~Furuta and T.~Inami, Mod. Phys. Lett. {\bf A15}
 (2000) 997.
\bibitem{NOS} C.~N\'u\~nez, K.~Olsen and R.~Schiappa,
JHEP {\bf 0007} (2000) 030.
\bibitem{GGR} H.O. Girotti, M. Gomes, V.O. Rivelles, A.J. da Silva,
hep-th/0005272.
\bibitem{AGB} L.~Alvarez Gaum\'e and J.F.L.~Barbon, hep-th/0006209.
\bibitem{AGW} L.~Alvarez Gaum\'e and S.R.~Wadia, hep-th/0006219.
\bibitem{MS2} E.~Moreno and F.A.~Schaposnik,
hep-th/0008118.
\bibitem{GP}A.M.~Ghezelbash and S.~Parvizi, hep-th/0008120.
\bibitem{GM} J.M.~Garc\'\i a-Bond\'\i a and C.P.~Mart\'\i n,
\bibitem{redlich} A.N. Redlich, Phys.Rev.Lett. {\bf 52} (1984) 18.
\bibitem{grs} R.E.~Gamboa Sarav\'\i , G.~Rossini and F.A.~Schaposnik,
Int. Jour. Mod. Phys. {\bf A11} (1996) 2643.
\bibitem{CG} C.P.~Mart\'\i n and D.~S\'anchez-Ruiz, Phys.Rev.Lett.
{\bf 83} (1999)  476.
\bibitem{CG1} T.~Krajewski and R.~Wulkenhaar, Int.J.Mod.Phys.
{\bf A15} (2000) 1011.
\bibitem{Wi} E.~Witten, Comm. Math. Phys. {\bf 121} (1989) 351.
\bibitem{MSe} G.~Moore and N.~Seiberg, Phys. Lett. {\bf 220} (1989)
422.
\bibitem{Eal} S.~Elitzur, G.~Moore, A.~Schwimmer and N.~Seiberg,
Nucl. Phys. {\bf B326} (1989) 108.

\end{thebibliography}
\end{document}